\documentclass[
    ,final            % use final for the camera ready runs
  ,numberedheadings % uncomment this option for numbered sections
  ]
  {aipproc}

\usepackage{amsmath,amssymb,graphicx}
\layoutstyle{8x11single}
\newcommand{\xoeff}{\tilde X_0}
\newcommand{\xomem}{\tilde X_0^\mathrm{\tiny MEM}}

\begin{document}

\title{Towards flavored bound states beyond rainbows and ladders}


\author{B. El-Bennich}{address={Laborat\'orio de F\'isica Te\'orica e Computacional, Universidade Cruzeiro do Sul, S\~ao Paulo 01506-000 SP, Brazil}}
\author{E. Rojas}{address={Laborat\'orio de F\'isica Te\'orica e Computacional, Universidade Cruzeiro do Sul, S\~ao Paulo 01506-000 SP, Brazil}}
\author{M.~A. Paracha}{address={Laborat\'orio de F\'isica Te\'orica e Computacional, Universidade Cruzeiro do Sul, S\~ao Paulo 01506-000 SP, Brazil},
altaddress={Centre for Advanced Mathematics and Physics, National University of Science and Technology, Islamabad, Pakistan}}
\author{J.~P.~B.~C. de Melo}{
  address={Laborat\'orio de F\'isica Te\'orica e Computacional, Universidade Cruzeiro do Sul, S\~ao Paulo 01506-000 SP, Brazil} }

\begin{abstract}
 We give a snapshot of recent progress in solving the Dyson-Schwinger equation with a beyond rainbow-ladder ansatz for the dressed quark-gluon 
 vertex which includes ghost contributions. We discuss the motivations for this approach with regard to heavy-flavored bound states and form factors 
 and briefly describe future steps to be taken.
\end{abstract}

\maketitle

%%%%%%%%%%%%%%%%%%%%%%%%%%%%%%%%%%%%%%%%%%%%
%% MAINMATTER
%%%%%%%%%%%%%%%%%%%%%%%%%%%%%%%%%%%%%%%%%%%%

\section{Motivation \label{motiv}}

The challenge of understanding bound states in terms of the elementary fields of a given quantum field theory is persistent
and of particular interest in Quantum Chromodynamics (QCD). The question of how precisely the quarks and gluons form
hadrons immediately leads into the domain of relativistic quantum fields whose key properties can only be understood
with nonperturbative methods. Paramount among the challenges is the understanding of confinement and dynamical chiral 
symmetry breaking (DCSB), both of which are likely to be intimately related~\cite{Bashir:2012fs}. 

Whereas certain simple two-body bound states and their resonances can be adequately described by potential models, this is not the 
case in QCD. Did light quarks not exist, the picture of string-like potentials arising from a flux tube between two (infinitely) heavy quarks 
would be correct. Yet, in the real world, where light current-quarks are ubiquitous, it is a feature of QCD that light-pair creation and 
annihilation effects are essentially nonperturbative and cannot be described by a quantum mechanical potential~\cite{Bashir:2012fs,Chang:2009ae}. 
Such potentials are a poor guide to understanding the Goldstone boson of QCD and must be necessarily fine-tuned. In particular, 
in typical applications to hadronic form factors with ``light" constituent quarks, the quark's propagation, $S(k) = (\gamma\cdot k-m_q)^{-1}$, 
is scale independent and does not describe confinement. In applications of relativistic quark models it  was noted that this can 
lead to significant model dependance at larger momentum transfers~\cite{ElBennich:2008qa,ElBennich:2008xy,daSilva:2012gf,ElBennich:2012ij,deMelo:2005cy}.  

Heavy-light mesons are of additional interest since they exhibit some features of light-quark confinement. The important 
asymmetry in quark masses of flavor-nonsinglet $Q\bar q$ mesons leads to a disparate array of energy scales to be dealt with 
in solving the meson's relativistic bound-state equation. Thus, heavy mesons provide an excellent opportunity to study 
additional aspects of nonperturbative QCD and can be used to test simultaneously all manifestations of the Standard Model, 
namely the interplay between electroweak and strong interactions. Some of the major advances in heavy quark effective theory 
(HQET)~\cite{Buchalla:1995vs} deal with factorization theorems allowing for a disentanglement of short-distance or hard physics, 
which includes electroweak interactions and perturbative QCD (pQCD) contributions, from long-distance or soft physics, dominated by 
nonperturbative hadronic effects. The systematic reorganization of weak and QCD interactions in HQET has been treated 
in various approaches; e.g., with QCD factorization (QCDF)~\cite{Beneke:1999br}, pQCD~\cite{Keum:2000wi} 
and soft-collinear effective theory (SCET)~\cite{Bauer:2000yr}. 

On the other hand, progress on nonperturbative matrix elements involving heavy-light states with flavor quantum numbers,
$C=\pm 1$ and $B=\pm1$, has been slower: while factorization theorems provide the means to systematically integrate out 
energy scales in the perturbative domain, valid in the infinitely heavy-quark limit, a reliable evaluation of the latter is notoriously 
difficult. Consider, for instance, the weak non-leptonic decay of a $B$ meson: $B\to M_1 M_2$. If $M_1$ is a heavy 
or light(er) meson and $M_2$ a light meson~\cite{Beneke:1999br}, then the decay amplitude can be schematically written as,
\begin{equation}
 \langle M_1 M_2 | O_i | B \rangle\  = \   \langle M_1| j_1 | B \rangle \langle M_2 | j_2 | 0 \rangle 
     \left [ 1 + \sum_n r_n \alpha_s^n + \mathcal{O} ( \Lambda_\mathrm{QCD}/m_h) \right ]  ,
\label{QCDfac}
\end{equation}
where $j_1$ and $j_2$ are the bilinear currents and $m_h$ is the heavy quark mass. The dimension-six effective four-quark operators, 
$O_i$, result from integrating out the weak gauge bosons $W^\pm$ in the operator product expansion. Multiplied by the appropriate 
Cabbibo-Kobayashi-Maskawa (CKM) matrix elements and Wilson coefficients, $C_i(\mu)$, which encode perturbative QCD effects above 
the renormalisation point $\mu$, the sum of these operators forms the heavy-quark effective hamiltonian. Neglecting power corrections 
in $\alpha_s$ and taking the limit $m_b\to \infty$, the naive factorization is recovered. Higher orders in $\alpha_s$ break the factorization, 
yet in the limit $m_h \gg \Lambda_\mathrm{QCD}$, pQCD corrections beyond the naive factorization can systematically be accounted 
for. In the case of $B$ decays, the factorization, formally suppressed in $\Lambda_\mathrm{QCD}/m_b$, can be broken by weak 
annihilation decay amplitudes~\cite{Zanetti:2010wj,Natale:2008wt} and final-state interactions between daughter hadrons 
\cite{ElBennich:2006yi,Boito:2007jh,Boito:2008zk,ElBennich:2009da,Leitner:2010fq,ElBennich:2011gm}. 

Moreover, since the charm quark is neither a light nor really a heavy quark, HQET may not be the adequate guide to charm physics 
and $\Lambda_\mathrm{QCD}/m_c$ corrections are significant~\cite{ElBennich:2010ha,ElBennich:2012tp}. Whilst effective 
Lagrangians based on approximate SU(4) flavor-spin symmetries\footnote{i.e., models which implement SU(4)$_F$ flavor symmetry 
in their Lagrangian approaches yet break these symmetries with the empirically known hadron masses.} are successful, 
see for example Refs.~\cite{Xiao:2013yca,Molina:2009eb,Krein:2010vp,Tsushima:2011kh}, care should be taken when this symmetry 
is applied to the Lagrangian's effective couplings. As has been noted, SU(4)$_F$ relations underestimate, for example, the $D\rho D$ 
coupling by a factor of four to five~\cite{ElBennich:2011py,Can:2012tx}. In this context, it is also noteworthy that SU(3)$_F$ and 
heavy-quark spin symmetry breaking effects are by no means insignificant and are manifest in the decay constant ratios, $f_{D_s}/f_D$, 
$f_{B_s}/f_B$ as well as  $f_{D^*}/f_D$ and $f_{B^*}/f_B$; see, e.g., Section~4 in Ref.~\cite{ElBennich:2012tp} and similar observations 
in lattice-QCD computations~\cite{Becirevic:2012ti}.

With respect to the weak decay constants, our focus in Eq.~\eqref{QCDfac} is on the hadronic matrix elements, $\langle M_1| j_1 | B\rangle$ 
and $\langle M_2 | j_2 | 0 \rangle$. The latter represents the weak decay constant of $M_2$, which in the case of a pseudoscalar 
meson is given by,
\begin{equation}
  f_M P_\mu = \langle 0 | \bar q_a \gamma_\mu\gamma_5 q_b | M \rangle  \ =   \ Z_2 \int\!\! \frac{d^4k}{(2\pi)^4}\ \mathrm{tr_{CD}} \!
              \left [ \gamma_5 \gamma_\mu S_q^a(k+\eta P) \Gamma_M^{a,b}(k;P)  S_q^b (k- \hat{\eta} P) \right ] \ ,
    \label{decayconst}          
\end{equation}
where  $a,b$ collect flavor and color indices, $Z_2$ is the wave function renormalization constant, $S_q (k)$ are dressed quark 
propagators and $\Gamma (k;P)$ is the mesons's Bethe-Salpeter amplitude (BSA). Note that in a Poincar\'e invariant treatment, 
the BSA and the weak decay constant --- and any other hadronic matrix element --- are independent of the momentum partitioning 
parameters $\eta+\hat \eta =1$. 
In some cases, for instance for the pseudoscalar mesons, $\pi$, $K$, $D$, $D_s$ and $B$, the values of the weak decay constants are 
known experimentally~\cite{Beringer:1900zz}. Much effort has also been invested to determine the heavy-light decay constants with 
lattice-regularized QCD, recently with unquenched fermions and increasingly better extrapolations to the continuum limit and physical pion 
masses~\cite{Becirevic:2012ti,Dimopoulos:2011gx,Na:2012iu,Bazavov:2011aa,Namekawa:2011wt,Bazavov:2012dg,Carrasco:2013zta}. 
Note that while the $c$-quark is treated as a propagating mode, the $b$-quark is usually implemented as a static fermion in lattice calculations 
of $f_B$ and $f_{B_s}$.

The first form factor in Eq.~\eqref{QCDfac}, $ \langle M_1| j_1 | B \rangle$, describes the transition of a heavy to a light(er) meson via the weak $V-A$ 
current and includes the propagation of the light spectator quark. Their precise evaluation is crucial in determining branching fractions 
and associated $CP$-violating observables of non-leptonic weak $D$ and $B$ decays~\cite{ElBennich:2006yi,Boito:2008zk,ElBennich:2009da} and 
oscillations~\cite{Leitner:2010fq,ElBennich:2011gm}. In semi-leptonic decays of heavy and light mesons they play a pivotal role in the determination 
of the Standard Model parameters, more precisely its weak sector via the CKM matrix elements. For example, the $K_{e3}$ decay, 
$K\to\pi e\nu_{e}$, can be used to extract the matrix element $|V_{us}|$; $D_{e3}$ to obtain $|V_{cs}|$; and the semi-leptonic decays of $B$  
and $B_{c}$ mesons, in particular $B\to D\ell\nu_{\ell}$, $B_c\to D_s^{\ast}\ell^+\ell^-$  and $B\to \pi\ell\nu_{\ell}$, inform the matrix elements 
$V_{cb}, V_{cs}$ and $V_{ub}$, respectively~\cite{CR,AP}.
As an example, we consider the case of heavy $H (0^{-+})$ to light(er)  $P(0^{-+})$ transitions mediated by the weak HQET operators, 
$\bar q_l \gamma_\mu(1-\gamma_5) Q$, where the hadronic matrix element is completely described by two Lorentz vectors,
\begin{equation}
\langle  P(p_2) | \bar q_l \gamma_\mu (1 - \gamma_5 ) Q | H (p_1) \rangle \  =  \ F_+(q^2)  P_\mu   +   F_-(q^2) q_\mu  \ ,
\label{HPSformfac} 
\end{equation}
with the total heavy-meson momentum, $P_\mu=(p_1+p_2)_\mu$, $P^2=-M_H^2$, $q_\mu=(p_1-p_2)_\mu$, $Q=c,b$ and $q_l = u,d,s$. 

Applications of light-quark propagators solutions of QCD's Dyson-Schwinger equations (DSE) in conjunction with the heavy-quark expansion  
to the form factors, $F_+(q^2)$ and $F_-(q^2)$, are in qualitative and quantitative agreement with heavy-quark symmetry~\cite{Ivanov:1998ms,Ivanov:2007cw}. 
Yet, when the form factors in Eq.~\eqref{HPSformfac} are calculated both ways~\cite{Ivanov:1998ms,Ivanov:2007cw}, namely with the fully 
dressed heavy-quark propagator and the propagator in the heavy-quark limit, it is possible to verify the validity of said limit: corrections are of the order of 
$\simeq 20 - 30$\% are encountered in $b\to c$ transitions and can be as large as a factor of 2 in $c\to d$ transitions, as verified in a vast array of 
light- and heavy-meson observables~\cite{Ivanov:2007cw}. Moreover, the following ratios of transition form factors serve as a measure of 
SU(3)$_F$ breaking:
\begin{equation}
 \frac{F_+^{B\to K}(0)}{F_+^{B\to\pi}(0)} = 1.23 \ , \quad   \frac{A_0^{B\to K^*}(0)}{A_0^{B\to\rho}(0)} = 1.25 \ .
\label{formratio}
\end{equation}
In Eq.~\eqref{formratio}, $A_0(q^2)$ are the appropriate form factors in $H (0^{-+})$ to $V(1^{--})$ transitions~\cite{Ivanov:2007cw}. The flavor breaking is of 
similar order as for the decay constant ratios, $f_{D_s}/f_D$ and $f_{B_s}/f_B$, discussed in detail in Ref.~\cite{ElBennich:2012tp}.

For a summary of heavy-to-light transition form factor data from lattice-regularized QCD, see the review in Ref.~\cite{Aoki:2013ldr}; we merely 
stress that contemporary lattice results are obtained for large squared-momentum transfer, i.e., $q^2 \gtrsim 16$~GeV$^2$ in the case of $B\to \pi$ 
transitions and $q^2 \gtrsim 7$~GeV$^2$ or $q^2_{\mathrm{max}}$ in $B\to D$ transitions~\cite{de Divitiis:2007ui,Bailey:2012rr,Atoui:2013zza}.
Values at low $q^2$ must necessarily be extrapolated by means of appropriate parametrizations~\cite{Becirevic:1999kt}. The rather strong quantitative 
differences between several predictions for the $B\to \pi$ form factor are emphasized in table~1 of Ref.~\cite{ElBennich:2009vx} from which it is plain 
that model dependence is still the major obstacle to a precision calculation of even the simplest transition form factors. 

In order to significantly improve on these form factor predictions in nonperturbative continuum QCD approaches, much progress has to be made in the 
hadronic description of heavy-light bound states. Within the framework of the DSEs and Bethe-Salpeter equations (BSE), recent efforts come to the 
conclusion that the so-called rainbow-ladder truncation fails to adequately reproduce basic static observables, such as the weak decay constant via 
Eq.~\eqref{decayconst}~\cite{Nguyen:2010yh}. Other nonperturbative quantities are light-front distribution amplitudes which play an important role in 
QCDF analyses of hard exclusive processes. In particular, light-front projections of the pseudoscalar's BSA, namely the pseudoscalar and pseudotensor 
projections, are identified as twist-three two-particle distribution amplitudes for which estimations with QCD sum rules exist~\cite{Braun:1989iv,Ball:1998je,Ball:2006wn}. 
The pseudoscalar projection has recently been derived~\cite{Chang:2013epa} and the same method may be applied to the pseudotensor projection. 
These projections can then be extended to $D$ and $B$ mesons provided a reliable BSA exists which allows for a faithful reproduction of experimental 
data on the respective weak decay constants. It is the aim of this contribution to sketch the path to a successful computation of the heavy-light meson's BSA.

\section{Two, three, four ... how many points in a meson?}

In the continuum formulation, QCD's two-point Green functions are described by DSE, which provide the adequate nonperturbative approach. 
Likewise, mesons are quark-antiquark bound states which appear as poles in the 2-quark, 2-antiquark Green's function, 
$G^{(4)} = \langle 0 |q_1q_2  \bar q_1 \bar q_2 |0 \rangle$ These poles are found from studies of the inhomogeneous  
pseudoscalar and axialvector BSE~\cite{Bashir:2012fs,Roberts:2007ji,Maris:2005tt,Krassnigg:2003wy}, as will be
discussed shortly.

The Dyson or gap equation determines how quark propagation is influenced by interaction with the gauge fields. For a given
quark flavor, the solutions of the quark DSE,
\begin{equation}
   S^{-1}(p)  =   \, Z_2 (i\, \gamma\cdot p + m^{\mathrm{bm}}) \ +  \ Z_1\, g^2\! \!\int^\Lambda_k \!\!\ \Delta^{\mu\nu}(q) \frac{\lambda^a}{2} 
                    \gamma_\mu\, S(k) \,\Gamma^a_\nu (-p,k,q) \ ,
\label{DSEquark}
\end{equation}
where $\int_k^\Lambda\equiv \int^\Lambda d^4k/(2\pi)^4$ represents a Poincar\'e invariant regularization of the integral with the regularization mass 
scale, $\Lambda$, and $Z_{1,2}(\mu,\Lambda )$ are the vertex and quark wave-function renormalization constants. The (infinitely many) nonperturbative 
interactions alter the current-quark bare mass, $m^{\mathrm{bm}}(\Lambda)$, which receives corrections from the self-energy given by the second
term in Eq.~\eqref{DSEquark}, where the integral is over the dressed gluon propagator,  $\Delta_{\mu\nu}(q)$, the dressed quark-gluon vertex, 
$\Gamma^a_\nu (-p,k,q)$, and $\lambda^a$ are the usual SU(3) color matrices of the fundamental representation. The gluon propagator is purely 
transversal in Landau gauge, which offers advantages in phenomenological interaction ans\"atze~\cite{Bashir:2012fs,Raya:2013ina}:
\begin{equation}
   \Delta^{ab}_{\mu\nu} (q) =  \delta^{ab} \left( g_{\mu\nu} - \frac{q_\mu q_\nu}{q^2} \right) \Delta ( q^2 ) \ .
\end{equation}   
The quark-gluon vertex is given by $\Gamma^a_\mu (p_1, p_2, p_3) = g \, \frac{\lambda^a}{2} \, \Gamma_\mu (p_1, p_2, p_3) $
with  the convention: $p_1+p_2+p_3 =0$.

The solutions to the gap equation~(\ref{DSEquark}) are of the general form $S(p) =  \left [ i\, \gamma \cdot p\ A(p^2) + B(p^2) \right ]^{\!-1}$
with the renormalization condition, $Z (p^2) = 1/A (p^2)  |_{p^2 = \mu^2} = 1$  at large spacelike $\mu^2\gg \Lambda_\mathrm{QCD}^2$. 
The mass function,  $M(p^2)=B(p^2\!, \mu^2)/A(p^2\!, \mu^2)$, is independent of the renormalization point $\mu$. In order to make quantitative 
matching with pQCD, another renormalization condition,
\begin{equation}
\left.  S^{-1}(p) \right |_{p^2=\mu^2}  = \  i\  \gamma \cdot p \ + m(\mu )  \ ,
\label{massmu_ren}
\end{equation}
is imposed, where $m(\mu )$ is the renormalized running quark mass.

Before discussing Bethe-Salpeter amplitudes, we  turn our attention to the quark-gluon vertex $\Gamma_\mu (p_1, p_2, p_3)$, which is one 
of QCD's three-point functions and satisfies its own BSE. In perturbation theory, that is for momenta $p_1^2=p_2^2=p_3^2 \gtrsim \mu^2$,
quark dressing effects are suppressed and $\Gamma_\mu (p_1, p_2, p_3) \to \gamma_\mu$. However, since the tremendous impact of DCSB 
on $Z(p^2)$ and $M(p^2)$ is nowadays well established, it is natural to accept that this also be true for the corresponding three-point functions. 

In applications to hadron physics, practical models for the fermion-gauge boson vertex ought to satisfy fundamental symmetries of QCD. 
General ans\"atze to the nonperturbative vertex impose constraints of quantum field theory; as just mentioned, one insists that the vertex must 
reduce to the bare vertex $\gamma_\mu$ in the large-momentum limit (when dressed propagators can be replaced by bare propagators); it must 
have the same transformation properties as the bare vertex under charge conjugation $C$, parity transformation $P$ and time reversal $T$; 
it must ensure gauge covariance and invariance; and one demands that the vertex must be free of kinematic singularities. Finally, the 
full nonperturbative vertex can always be decomposed into a longitudinal and a transverse part, $\Gamma_\mu (p_1, p_2, p_3) = 
\Gamma^\mathrm{L}_\mu (p_1, p_2, p_3)  +  \Gamma^\mathrm{T}_\mu (p_1, p_2, p_3)$~\cite{Ball:1980ay}. 
Clearly, gauge invariance is not satisfied for a bare vertex since it does not satisfy the Ward-Green-Takahashi identity (WGTI),
\begin{equation}
  i\, \gamma\cdot p_3\ \neq \  -i \gamma \cdot\!p_1\ A(p_1^2) +B(p_1^2) - i \gamma \cdot\!p_2\ A(p_2^2) - B(p_2^2) \ .
\end{equation}
Models that are largely consistent with the field theoretical constraints just mentioned have also been used to represent the 
dressed quark-gluon vertex, the most prominent amongst which is the Ball-Chiu ansatz for the longitudinal vertex~\cite{Ball:1980ay}. However, 
while employed in studies of hadronic observables, the Ball-Chiu vertex satisfies a WGTI whereas the true quark-gluon vertex satisfies a
Slavnov-Taylor identity (STI). The form of the latter, see Eq.~\eqref{STI}, makes it plausible that within certain approximations a solution of
\begin{equation}
    p_{3\mu} i\,  \Gamma_\mu (p_1,p_2,p_3) = \mathcal{B}(p_3^2) \left [ S^{-1} (-p_1) -  S^{-1} (p_2) \right ]
 \label{STIsol}   
\end{equation}
can provide a reasonable approximation to the correct vertex. 

In view of the scarce information on the quark-gluon vertex from first principle calculations, the strategy to combine different nonperturbative 
approaches to QCD was explored in Ref.~\cite{Rojas:2013tza}. Therein, lattice-QCD data for the dressed-quark functions, $A(p^2)$ and 
$B(p^2)$~\cite{Bowman:2005vx,Furui:2006ks}, as well as for the gluon and ghost propagators, $\Delta(q^2)$ and $F(q^2)$~\cite{Bogolubsky:2009dc,Oliveira:2012eh}, 
were employed to numerically extract a momentum-dependent effective function $\xoeff (q^2)$ from the quark gap equation via an inversion
procedure. In order to apply this inversion, one defines a "ghost-improved" Ball-Chiu vertex~\cite{Aguilar:2010cn,Fischer:2003rp},
\begin{equation}
\tilde \Gamma^\mathrm{BC}_\mu (p_1,p_2,p_3 ) = \xoeff (p^2_3)\,  F(p^2_3)\, \Gamma^\mathrm{BC}_\mu (p_1,p_2,p_3) \ ,
\label{xovertex}
\end{equation}
which can be derived from the constraints of the STI,
\begin{equation}
   p_{3\mu} \ i\, \Gamma_\mu ( p_1, p_2, p_3 )  =  F(p^2_3) \Big [ S^{-1} ( -p_1) \, H( p_1, p_2, p_3 ) 
    -    \overline H (p_2, p_1, p_3) S^{-1}(p_2) \Big ] \ ,
\label{STI}
\end{equation}
where $F(q^2)$  is the ghost-dressing function and the quark-ghost scattering kernel is parameterized in terms of the matrix-valued function, 
$H( p_1, p_2, p_3 )$, and its conjugate, $\overline H ( p_1, p_2, p_3 )$~\cite{Davydychev:2000rt}. The decomposition of these two functions 
in terms of Lorentz covariants requires eight form factors,
\begin{eqnarray}
  H( p_1, p_2, p_3 ) & = &  X_0\, \mathbb{I}_D + i\, X_1 \, \gamma\cdot p_1 + i\, X_2 \, \gamma\cdot p_2 + i\, X_3 \, \sigma_{\alpha\beta} 
  p^\alpha_1 p^\beta_2\, ,  \\
 \overline H( p_2, p_1, p_3 ) & = & \overline X_0\, \mathbb{I}_D - i\, \overline X_2\, \gamma\cdot p_1 - i\,\overline X_1\, \gamma\cdot p_2 +
    i\,  \overline X_3 \, \sigma_{\alpha\beta} p^\alpha_1 p^\beta_2\,  .
 \hspace{6mm}
\end{eqnarray}
Perturbative expressions for the form factors $X_i$ have been computed to one-loop order~\cite{Davydychev:2000rt} and yield 
$X_0 = 1 + \mathcal{O}(g^2)$ and $X_i = \mathcal{O}(g^2)$, $i = 1, 2, 3$. Thus, $X_0$ is the dominant form factor at large momenta 
and using the approximations $X_{1,2,3} \simeq 0$  and $\overline X_0  = X_0 = \xoeff (q^2)$ the dressed quark-gluon vertex reduces 
to the expression in Eq.~\eqref{xovertex} which satisfies the identity~\eqref{STIsol} with $\mathcal{B}(q^2) = \xoeff (q^2) F(q^2)$.

\begin{figure}[t]
\begin{minipage}{0.5\textwidth}
\vspace*{-1cm}
 \includegraphics[height=.31\textheight]{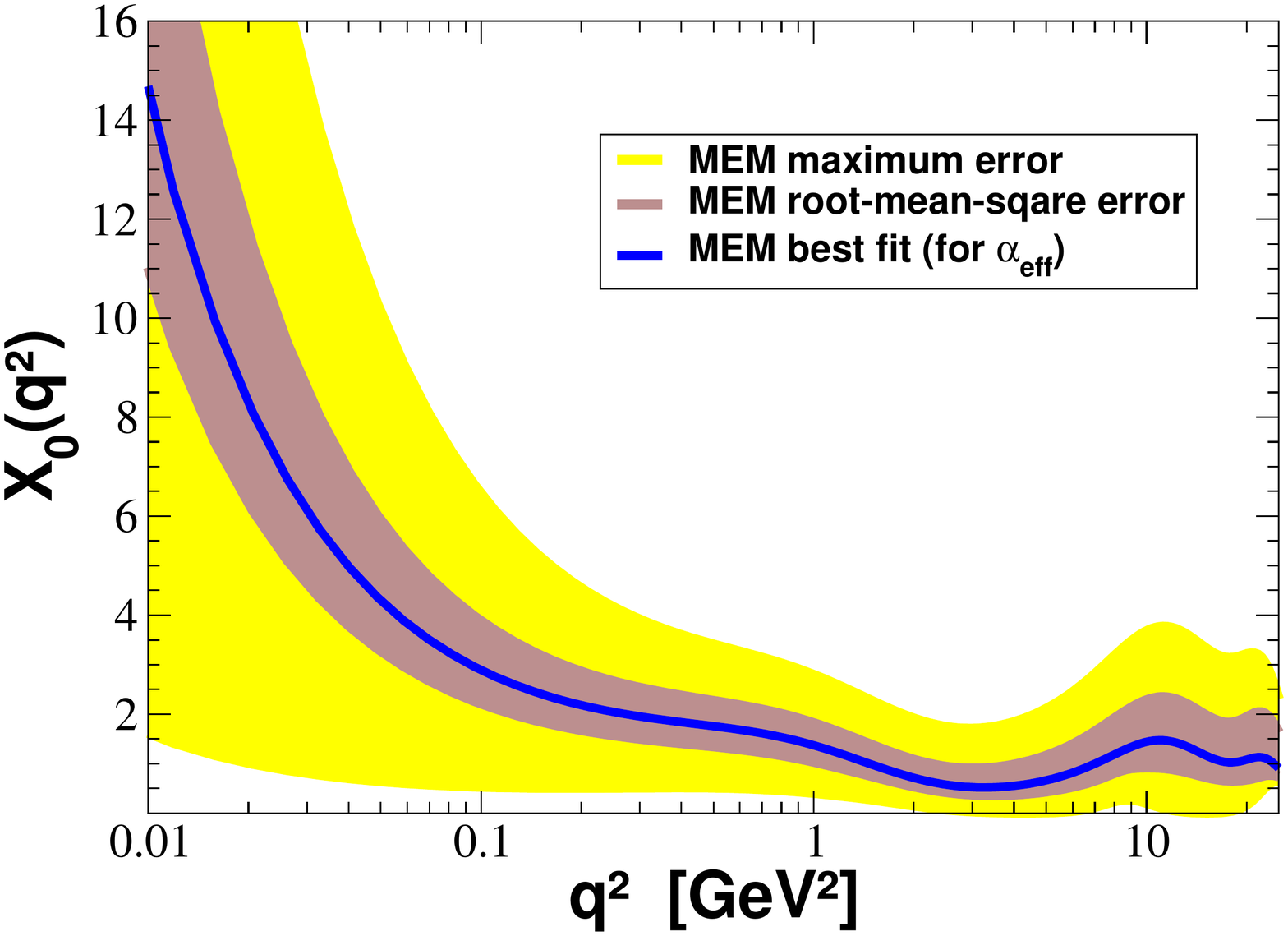}
\end{minipage}  
 \begin{minipage}{0.5\textwidth}
 \vspace*{-1.2cm}
  \includegraphics[height=.31\textheight]{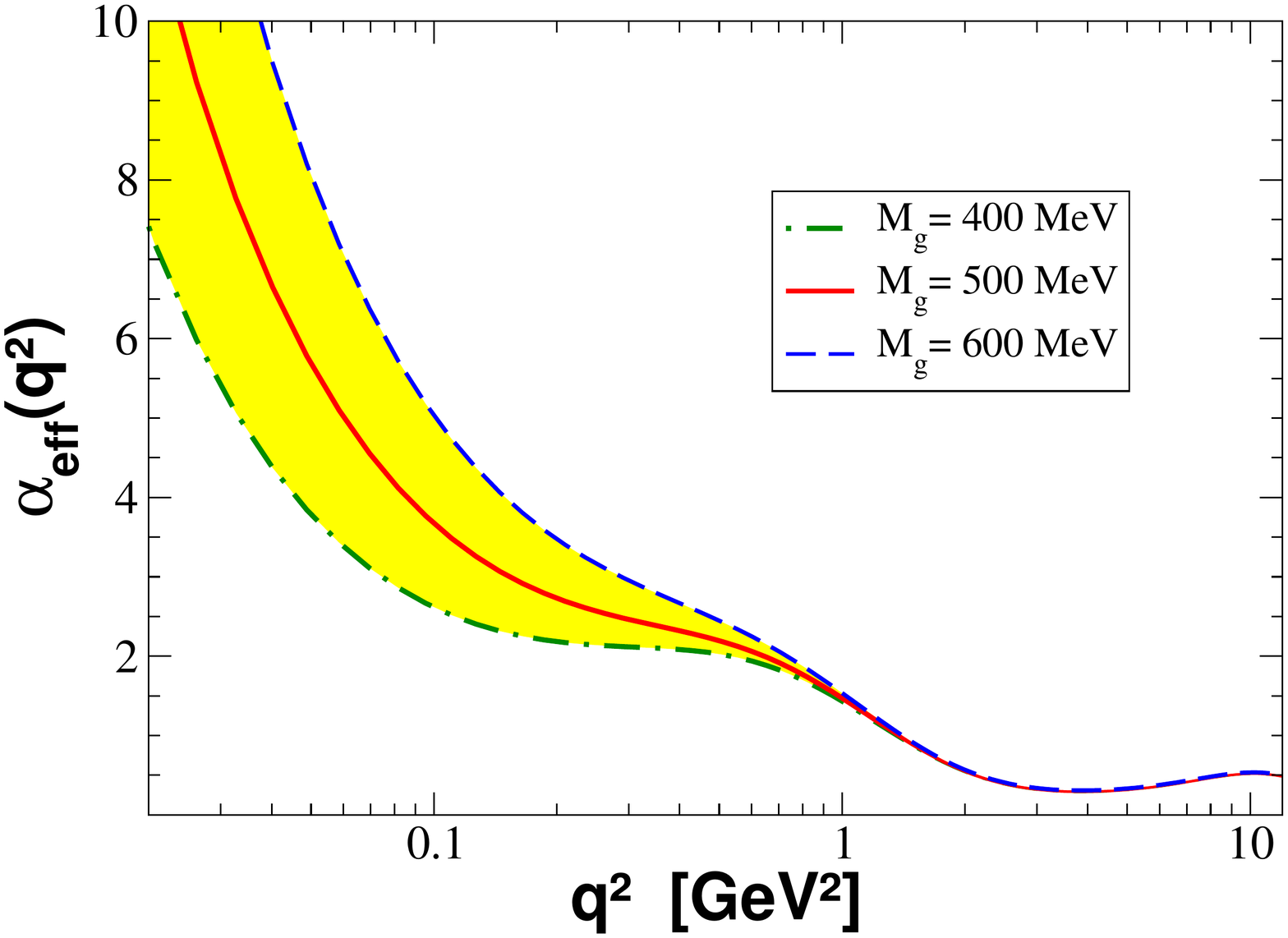}
   \vspace*{-7mm}
\end{minipage}  
 \caption{Left panel: the effective quark-gluon vertex function, $\xomem$, from a nonlinear inversion based on MEM;  see Ref.~\cite{Rojas:2013tza} 
 for details. Note that the functional form of $\xomem$ differs from that in Ref.~\cite{Rojas:2013tza} where  $\xomem (q^2 \to 0) \to 1$. Here it is clearly 
 enhanced in the infrared due to the requirement that $q^2 \xoeff (q^2)$ be finite in the MEM inversion. However, considering the case of maximal correlation 
 of the MEM fit parameters, represented by the maximum error (yellow) band, the solution for $\xoeff$ in Ref.~\cite{Rojas:2013tza} is compatible with 
 the present one. Owing to the lack of lattice-QCD data on $A(p^2)$ and $B(p^2)$, the inversion procedure is simply not constrained below $0.14$~GeV. 
 Nonetheless, either solution yields the mass functions in the left-hand plot of Fig.~\ref{mxo} since the DSE kernel is vanishing for $q \lesssim 0.2$~GeV 
 and the functional form of $\xoeff$ thus becomes irrelevant in this momentum range.  
 Right panel: the effective charge defined in Eq.~\eqref{xovertex} with $\xoeff \equiv \tilde X_0^{\mathrm{MEM}}$. \vspace*{-2mm} }
\label{xofig}
\end{figure}

The quark-gluon vertex built from the resulting $\xoeff$ is enhanced in the infrared region and recovers the perturbative behavior as one 
approaches larger momenta. The two different inversion methods employed in Ref.~\cite{Rojas:2013tza}, linear regularization and the 
maximum entropy method (MEM), produce $\xoeff$ form factors compatible with each other for the range of momenta where lattice-simulation 
data is available, i.e. in the domain $\simeq 0.3-4$~GeV. For momenta in the range $0.3-1$~GeV, both regularizations feature a strong 
yet somewhat different enhancement of the generalized Ball-Chiu vertex, as depicted in Fig.~\ref{xofig} for the MEM case, which generates 
the DCSB observed in the lattice-QCD mass functions~\cite{Bowman:2005vx,Furui:2006ks}. As can be read from the figure, for 
momentum values above 4~GeV, the extraction of $\xoeff (q^2)$ becomes less reliable owing to the lack of lattice-data constraints. Similarly, the 
steep increase of $\xoeff (q^2)$ below $\simeq 0.1$~GeV$^2$ is to be taken with caution. Nonetheless, the functional form of $\xoeff$ above 
$q^2 \gtrsim 0.1$~GeV$^2$ bears strong similarities with that of the ghost-dressing function~\cite{Bogolubsky:2009dc,Oliveira:2012eh}, $F(q^2)$, 
which {\em a posteriori\/} justifies the prescription $F(q^2) \to F^2(q^2)$ employed in Ref.~\cite{Aguilar:2010cn}.
\begin{figure}[b]
\begin{minipage}{1\textwidth}
\vspace*{-4mm}
 \includegraphics[height=.28\textheight]{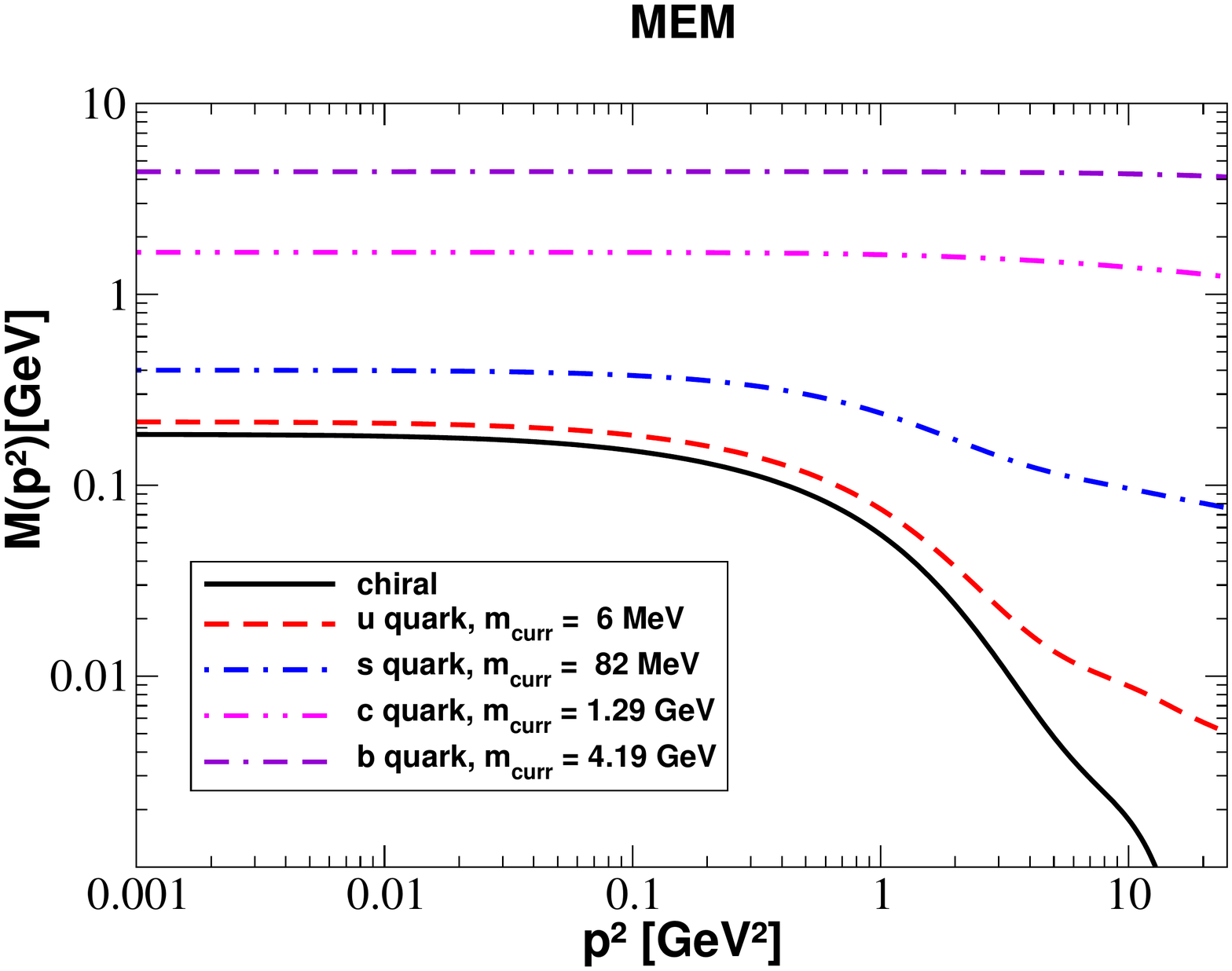}
\hfill \hspace*{3mm}
\includegraphics[height=.28\textheight]{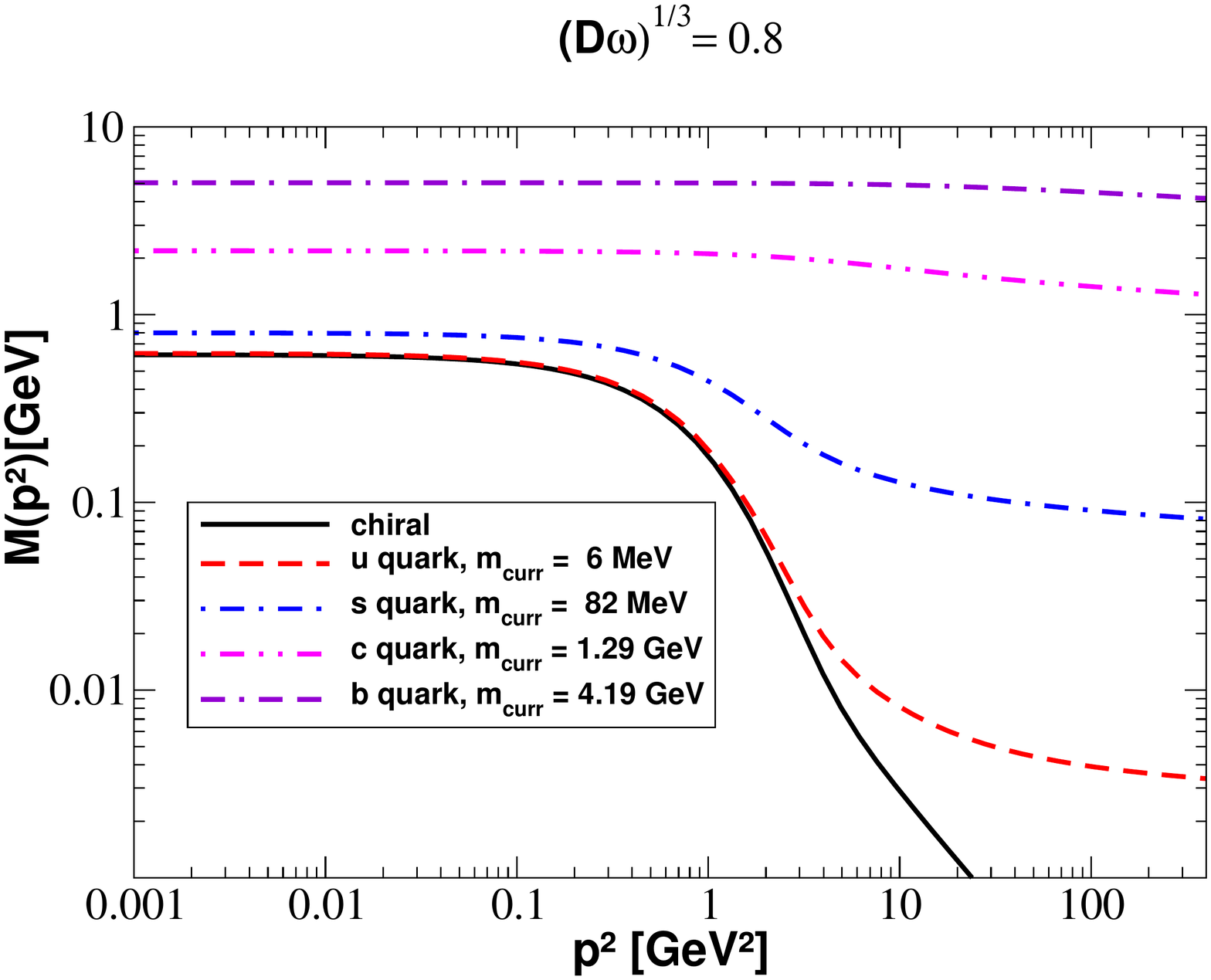}
\vspace*{-6mm}
 \caption{Flavor dependence of the solutions for the mass function $M(p^2) = B(p^2)/A(p^2)$; left panel: using the effective interaction model 
 of Eq.~\eqref{effalpha}; right panel: DSE with interaction model and given parameter set of Ref.~\cite{Qin:2011dd}.
\vspace*{-6mm}}
\end{minipage} 
\label{mxo}
\end{figure}
We thus, in analogy with Ref.~\cite{Ayala:2012pb}, define an effective charge via the combination,
\begin{equation}
 \alpha_{\mathrm{eff}} (q^2) = \alpha_s\ \xoeff (q^2) F (q^2) \Delta (q^2) \left [q^2 + m_g^2(q^2) \right ] \ ; \quad m_g^2 (q^2) = \frac{M_g^4}{q^2+M_g^2}  \ ,
\label{effalpha}
\end{equation}
plotted in Fig.~\ref{xofig}, where $\alpha_s (4.3\,\mathrm{GeV})=0.295$ and $M_g^2$ typically of the order $500-600$~MeV.

The flavor dependence of the solutions for the mass function is depicted in the left panel of Fig.~\ref{mxo}. For the light quarks, $u$ and $d$, 
the DCSB leads to $M(0) \simeq 220$~MeV in accordance with lattice results~\cite{Bowman:2005vx,Furui:2006ks}. Yet, in comparison with 
the best available phenomenological interaction model~\cite{Qin:2011dd} whose functional behavior accords qualitatively with results 
of modern DSE and lattice studies, the effect of the DCSB is much weaker: as seen in the right panel of Fig.~\ref{mxo}, the model of Ref.~\cite{Qin:2011dd} 
yields $M(0) \simeq 600$~MeV. Although the consequences of DCSB  are less marked for the heavy quarks' mass functions, which remain 
almost constant over a large momentum domain, we do note a difference $\Delta M_b (0) \simeq 600$~MeV between both models. It is thus expected 
that the interaction defined by Eq.~\eqref{effalpha} is too weak for applications to hadron phenomenology. This shortcoming may be remedied by the 
inclusion of the transverse quark-gluon vertex component which describes the quark's anomalous chromomagnetic moment~\cite{Chang:2010hb}.

\begin{figure}[t!]
\begin{minipage}{0.9\textwidth}
 \includegraphics[height=.24\textheight]{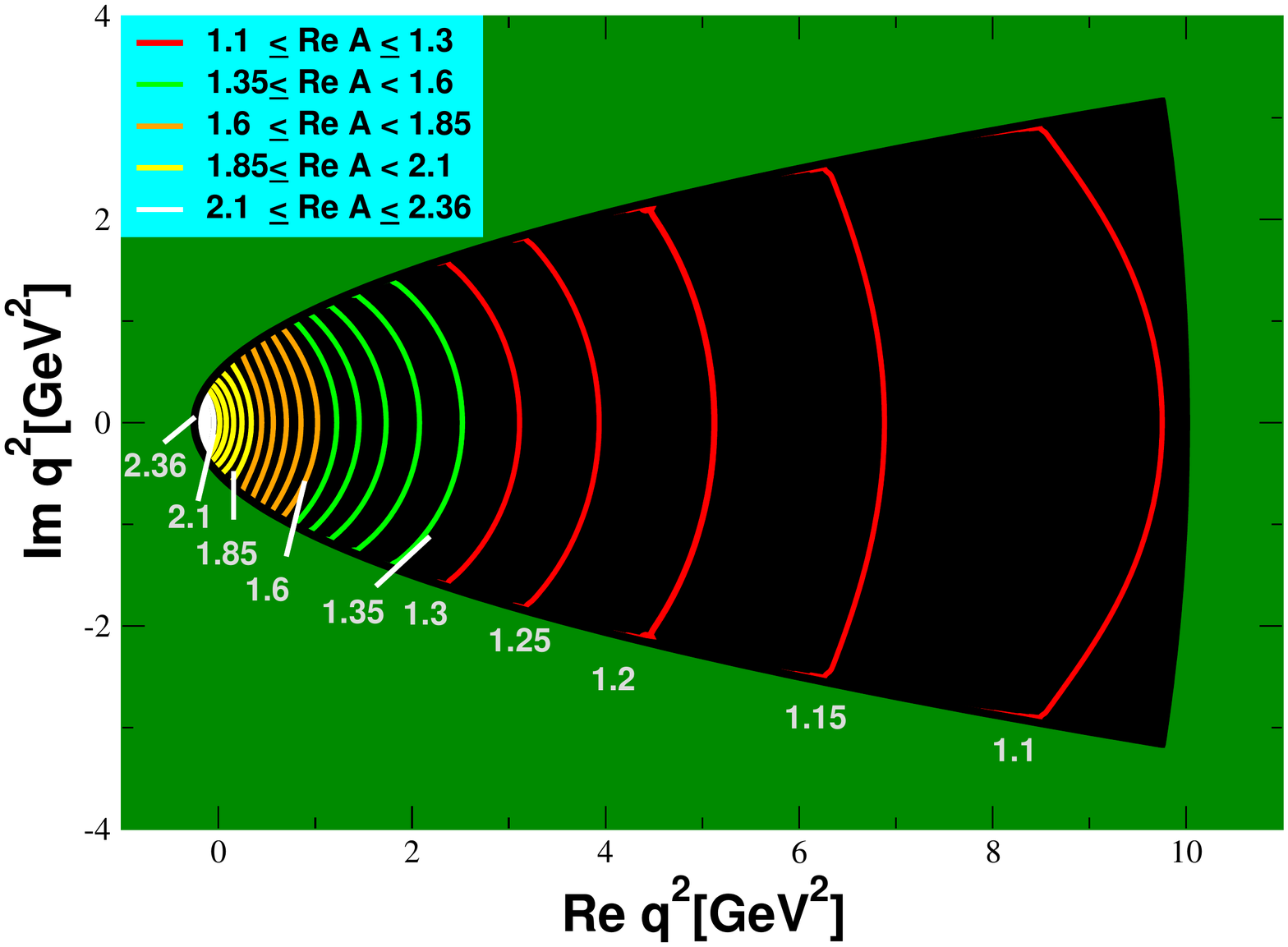}
 \hfill \hspace*{7mm}
\includegraphics[height=.24\textheight]{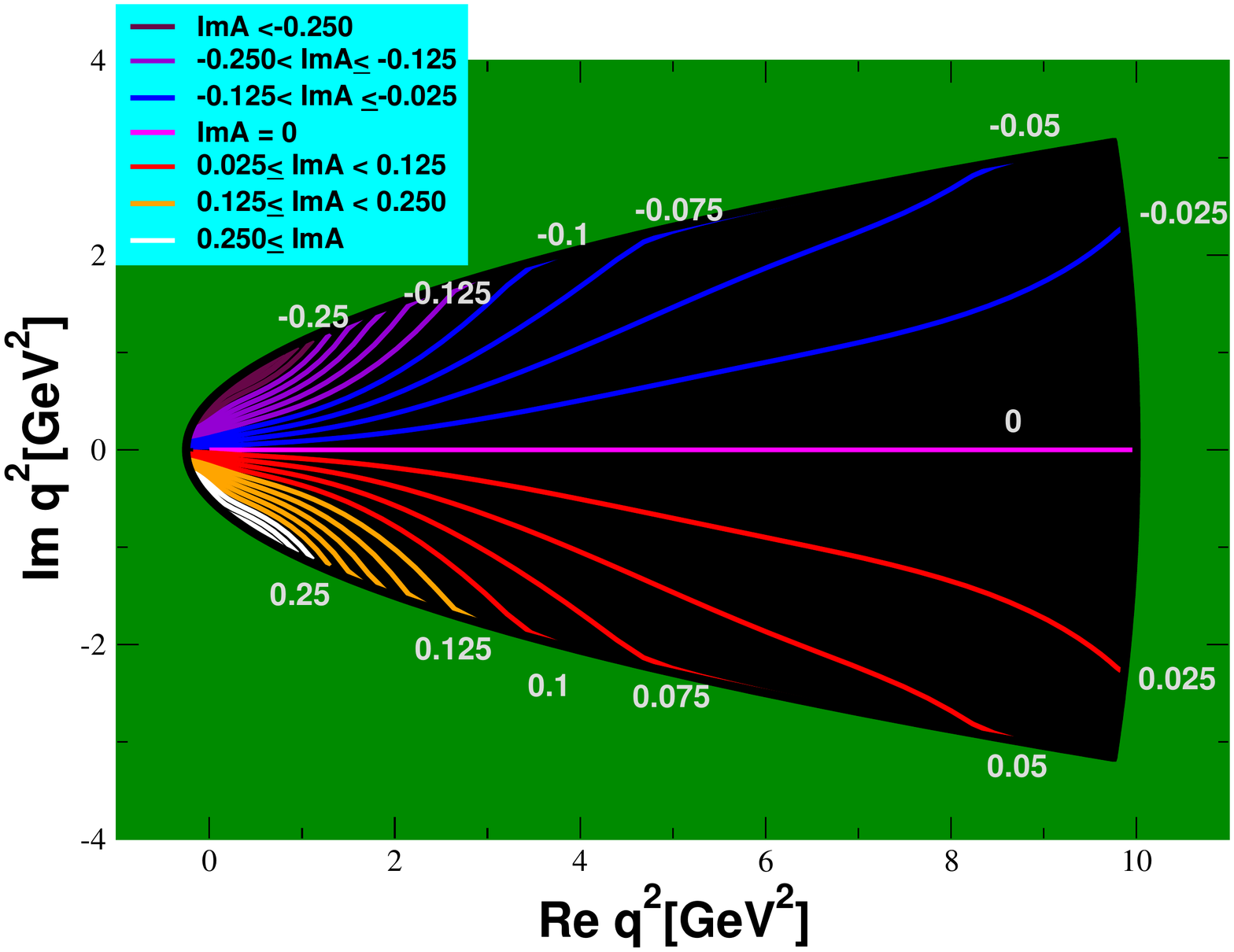}  
\vspace*{-8mm}
\end{minipage}
\end{figure}
\begin{figure}[t]
\begin{minipage}{0.9\textwidth}
 \includegraphics[height=.24\textheight]{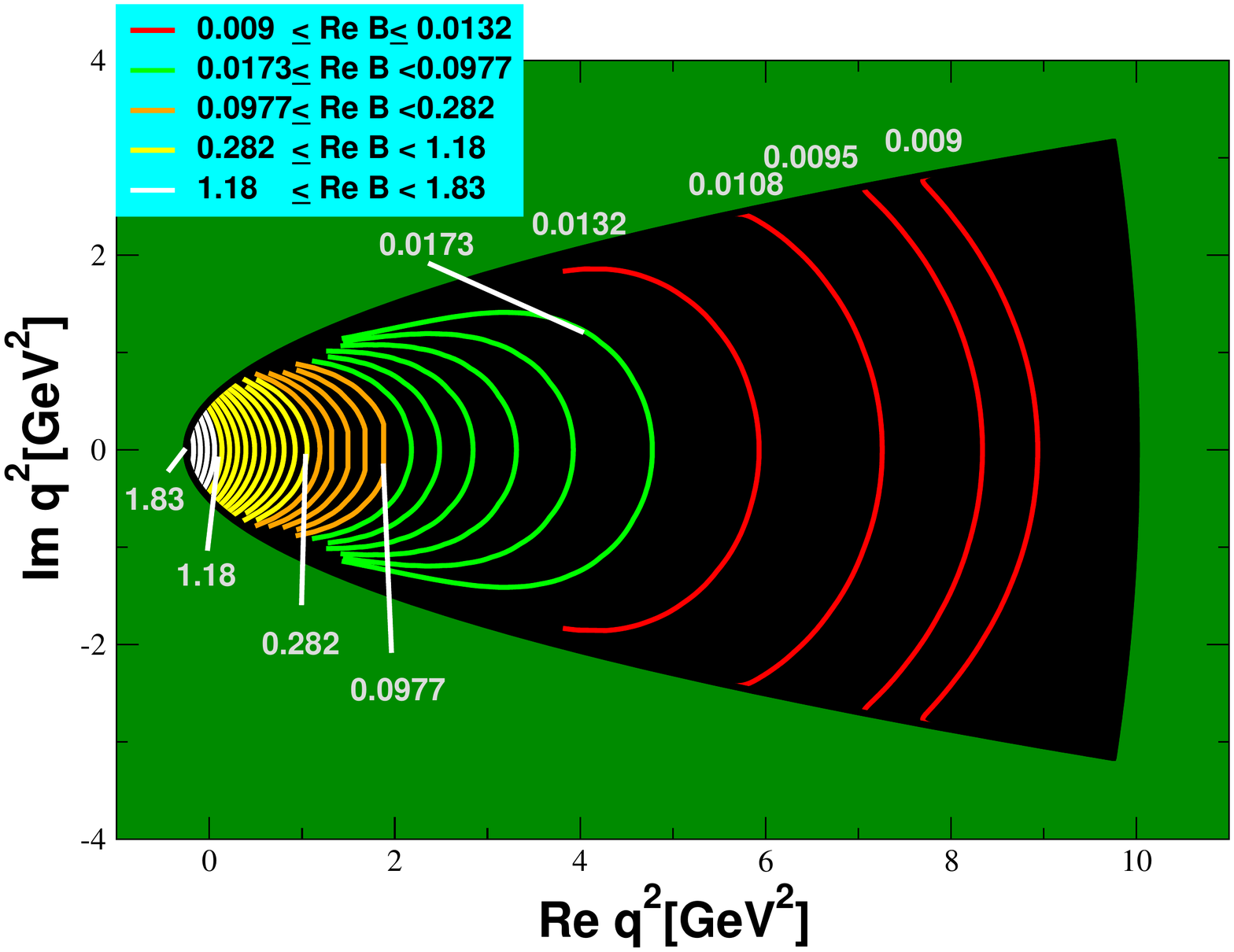}  
 \hfill  \hspace*{5mm} 
\includegraphics[height=.24\textheight]{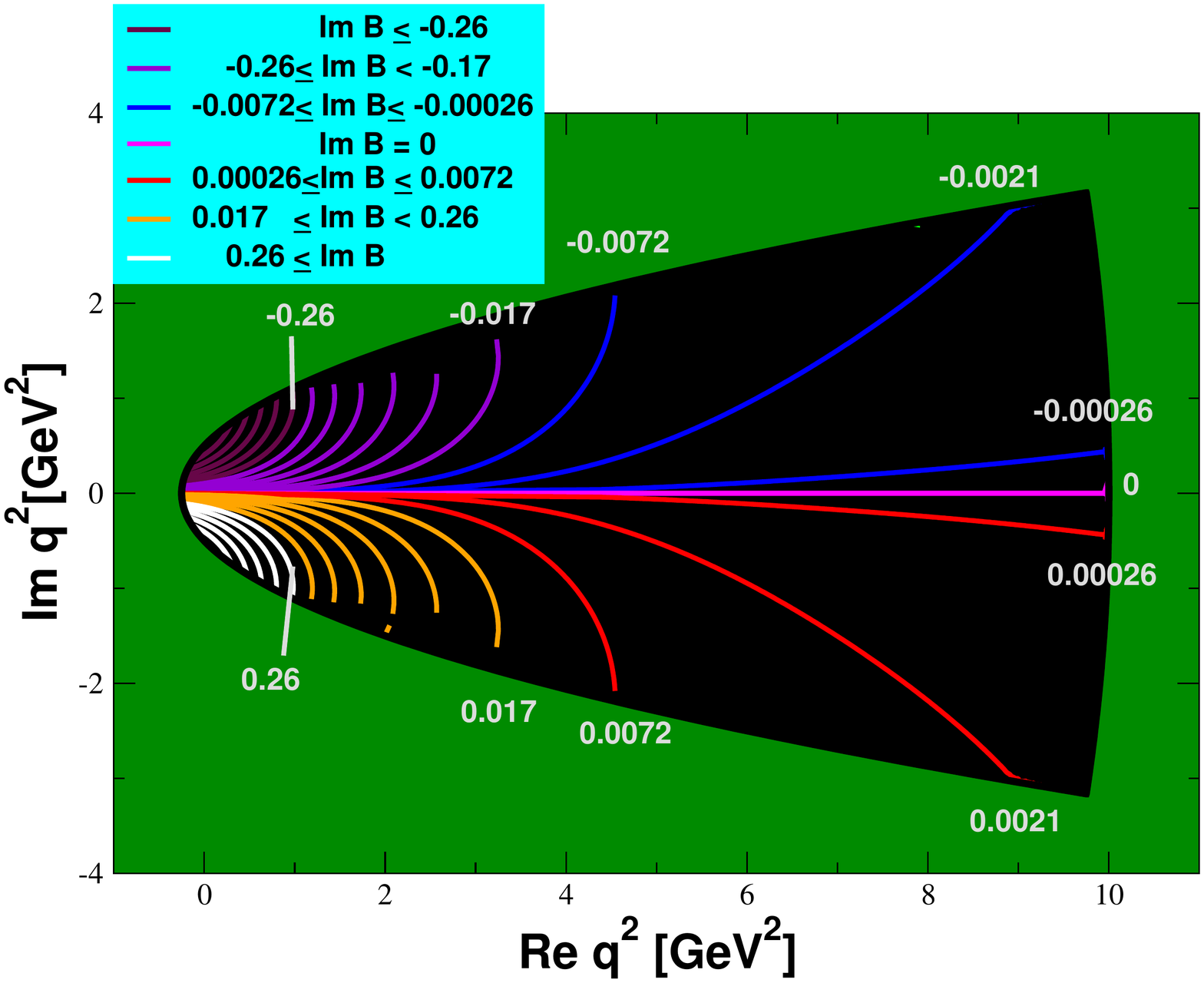}
\end{minipage}
\caption{Contour levels of the real and imaginary parts of the DSE solutions, $A(p^2)$ and $B(p^2)$, for complex momenta using the interaction model of Ref.~\cite{Qin:2011dd}.
\vspace{-2mm}}
\label{complex}
\end{figure}

Mass functions are not physical observables and to test the validity and efficacy of an interaction model it must stand the comparison with experimental data. 
To this end, bound state equations must be solved and while the propagator satisfies the gap equation, the vertices are determined by an inhomogeneous 
BSE. Consider, for instance, the exact inhomogeneous axialvector BSE~\cite{Chang:2009zb} which is valid when the quark-gluon vertex is fully dressed,
i.e. for an ansatz beyond the rainbow-ladder truncation:
\begin{eqnarray}
\label{BSE}
 \Gamma^{fg}_{5\mu}(k;P) & = &  Z_2\, \gamma_5 \gamma_\mu - g^2\! \int_q^\Lambda\!\! D^{\alpha\beta}Ê(k-q) \frac{\lambda^a}{2} \gamma_\alpha 
\ S_f(q_+) \,\Gamma^{fg}_{5\mu}(q;P) S_g(q_-) \frac{\lambda^a}{2} \Gamma^g_\beta (q_-,k_-) \nonumber \\
  & + &   g^2\!   \int_q^\Lambda\!\!\! D^{\alpha\beta}Ê(k-q) \frac{\lambda^a}{2} \gamma_\alpha\, S_f(q_+) \frac{\lambda^a}{2} \Lambda^{fg}_{5\mu\beta}(k,q;P)\ ,
\end{eqnarray}
where $P$ is the total meson momentum, $q_\pm=q\pm P/2, k_\pm = k\pm P/2$ and $f,g$ denote the flavor indices of a light-light or heavy-light bound state.
The 4-point Schwinger function $\Lambda^{fg}_{5\mu\beta}$ is entirely defined via the quark self energy. This comes about that a WGTI can be derived for 
the Bethe-Salpeter kernel whose solution provides a symmetry-preserving closed system of gap and vertex equations
\cite{Chang:2009zb,Chang:2011ei}. The pseudoscalar vertex, $\Gamma_5^{fg}(k;P)$, satisfies an analogous equation to Eq.~\eqref{BSE} and it is 
a well known feature of QCD that both the axialvector and pseudoscalar vertices exhibit poles whenever $P ^2 = - m^2_{M_n}$, where $m_{M_n}$
is the mass of the meson $M$ or any of its radial excitations~\cite{Maris:1997hd,Maris:1997tm} :
\begin{equation}
   \Gamma_{5\mu}(k;P) \big |_{P^2 +m_{M_n}^2 \simeq 0} \ = \ \frac{f_{M_n}\, P_\mu}{P^2+m_{M_n}^2} \Gamma_{M_n} (k;P) \ + \ \Gamma^{\mathrm{\,reg.}}_{5\mu} (k;P) \ ,
\end{equation}
Here, $ \Gamma_{M_n} (k;P)$ is the pseudoscalar bound state's BSA. The solutions of the BSE for $P^2 = -m_{M_n}^2$ in Euclidean momentum space requires 
the knowledge of the quark propagator at complex momenta whose squares lie inside a parabola and which in the past presented a considerable numerical challenge 
at large quark masses, $m_q > m_c$.  Improved numerical methods which facilitate the treatment of the quark's DSE are now available~\cite{Krassnigg:2009gd} and 
in Fig.~\ref{complex} we present the real and imaginary parts of the complex solutions $A(p^2)$ and $B(p^2)$, where $P^2 \simeq -1$~GeV$^2$. Solutions for
heavier quarks are currently being investigated and in conjunction with Eq.~\eqref{BSE} first results for the heavy meson's BSA using Eq.~\eqref{xovertex}
will soon be available.

\section{Epilogue} 

We have summarized recent progress towards computing the DSE and BSE for heavy-light systems beyond the rainbow-ladder truncation based
on an ansatz for the quark-gluon vertex which correlates the tensor structure of the Ball-Chiu vertex with a nonperturbative vertex function. The functional 
form of the latter is extracted from lattice-regulated QCD data on the quark's dressed propagator via an inversion of the DSE~\cite{Rojas:2013tza}. 
When the vertex model is re-inserted in the quark's DSE, its solutions yield mass functions which are qualitatively comparable with those obtained with
a recent interaction model~\cite{Chang:2010hb} but whose magnitude of DCSB is considerably smaller. Computations of the BSA for heavy-light systems
which make use of the exact form of the BSE, valid for the fully dressed quark-gluon vertex, and either interaction models  are underway. The first
test any beyond the rainbow-ladder ansatz for the quark-gluon vertex must pass is the numerical value one obtains for one of the most elementary observable, i.e. 
the weak decay constant which is more sensitive to the BSA normalization. We shall report results of its computation for $D_{(s)}$ and $B_{(s)}$ mesons in 
a future communication. This will be the first in a series of steps to obtain their form factors and parton distribution amplitudes that are our original motivation 
(see Section~\ref{motiv}) and for which the well known rainbow-ladder ans\"atze~\cite{Maris:1997hd,Maris:1997tm} are not phenomenologically valuable.

\begin{theacknowledgments}
This work is supported by the S\~ao Paulo Research Foundation, Funda\c{c}\~ao de Amparo \`a Pesquisa do Estado de S\~ao Paulo (FAPESP), and
the federal agency, Conselho Nacional de Desenvolvimento Cient\'ifico e Tecnol\'ogico (CNPq). We acknowledge valuable communication with 
Orlando Oliveira and Tobias Frederico. B.~E. would like to thank the organizers of the XXXVI {\em Reuni\~ao de Trabalho sobre F\'isica Nuclear no Brasil\/}
for their kind support.
\end{theacknowledgments}

%%%%%%%%%%%%%%%%%%%%%%%%%%%%%%%%%%%%%%%%%%%%%%%%
%% The bibliography can be prepared using the BibTeX program or
%% manually.
%%
%% The code below assumes that BibTeX is used.  If the bibliography is
%% produced without BibTeX comment out the following lines and see the
%% aipguide.pdf for further information.
%%
%% For your convenience a manually coded example is appended
%% after the \end{document}
%%%%%%%%%%%%%%%%%%%%%%%%%%%%%%%%%%%%%%%%%%%%%%%%

%%%%%%%%%%%%%%%%%%%%%%%%%%%%%%%%%%%%%%%%%%%%%%%%
%% You may have to change the BibTeX style below, depending on your
%% setup or preferences.
%%
%%
%% For The AIP proceedings layouts use either
%%%%%%%%%%%%%%%%%%%%%%%%%%%%%%%%%%%%%%%%%%%%

\bibliographystyle{aipproc}   % if natbib is available
%\bibliographystyle{aipprocl} % if natbib is missing

%\bibliography{sample}

\end{document}